\documentclass[reprint,prl,superscriptaddress,longbibliography]{revtex4-2}

\usepackage{amsmath}
\usepackage{graphicx,psfrag}
\usepackage[caption=false]{subfig}
\usepackage{verbatim}
\usepackage{color}
\usepackage{amsmath,amssymb}
\usepackage{listings}
\begin{document}
  
%\preprint{}

\title{Connecting lattice Boltzmann methods to physical reality by coarse-graining Molecular Dynamics simulations}


\author{Aleksandra Pachalieva}
\email[]{apachalieva@lanl.gov}
%\thanks{}
\affiliation{Computational Earth Science Group, Earth and Environmental Sciences Division, Los Alamos National Laboratory, Los Alamos, NM 87545, USA}
\affiliation{Center for Nonlinear Studies, Theoretical Division, Los Alamos National Laboratory, Los Alamos, NM 87545, USA}
\affiliation{Department of Mechanical Engineering, Technical University of Munich, 85748 Garching, Germany}

\author{Alexander J. Wagner}
\email[]{alexander.wagner@ndsu.edu}
%\thanks{}
\affiliation{Department of Physics, North Dakota State University, Fargo, ND 58108, USA}
\date{\today}

\begin{abstract}
The success of lattice Boltzmann methods has been attributed to their mesoscopic nature as a method derivable from a physically consistent microscopic model. Original lattice Boltzmann methods were Boltzmann averages of an underlying lattice gas. In the transition to modern lattice Boltzmann method, this link was broken, and the frequently used over\nobreakdash-relaxation to achieve high Reynolds numbers has been seen as lacking physical motivation.  While this approach has undeniable utility, it appeared to break the link to any underlying physical reality putting into question the special place of lattice Boltzmann methods among fluid simulation methods. In this letter, we show that over\nobreakdash-relaxation arises naturally from physical lattice gases derived as a coarse\nobreakdash-graining of molecular dynamics simulations thereby re\nobreakdash-affirming the firm foundation of lattice Boltzmann methods in physical reality.
\end{abstract}


\maketitle

Lattice Boltzmann methods originated from the revolutionary lattice gas method developed by Frisch, Hasslacher and Pomeau\,\cite{frisch_lattice-gas_1986}. In lattice gases, local collisions redistribute particles according to collision rules that conserve particle number and local momentum. The effectiveness of collisions can be tuned to some degree \cite{Frisch1987,Henon1987}, and the viscosity is reduced as a result. However, such collisions bring the simulations at most to a state of local equilibrium.

To derive the macroscopic behavior of these systems, one first takes a non\nobreakdash-equilibrium ensemble average of the lattice gas method, which results in a lattice Boltzmann equation\,\cite{Frisch1987}. This averaged lattice Boltzmann equation can be simulated directly, resulting in a noise\nobreakdash-free simulation method\,\cite{NamaraZanetti}. This increases computational efficiency, since additional averaging over lattice gas results is no longer required, which counteracts the higher computational cost arising from transitioning from a Boolean lattice gas to a lattice Boltzmann method requiring real numbers. Instead of directly averaging the lattice gas collision terms, as was done by McNamara \textit{et al.}\,\cite{NamaraZanetti}, one can relax the distributions towards local equilibrium distribution function\,\cite{higuera1989boltzmann}, resulting in the BGK approach.

The original lattice gas models were Boolean lattice gases, \textit{i.e.} only zero or one particle were allowed per occupation number, leading to Fermi\nobreakdash-Dirac, rather than Boltzmann equilibrium distribution\,\cite{Frisch1987}. This implied that the resulting hydrodynamic equations had mildly Galilean invariance violating terms\,\cite{Frisch1987}. Lattice Boltzmann methods, that abandoned their direct connection to the underlying lattice gas by imposing a Maxwell\nobreakdash-Boltzmann equilibrium distribution with the BGK collision term, removed those velocity dependent terms in the Navier\nobreakdash-Stokes equation\,\cite{qian1992lattice}. They are given by
\begin{equation}
    f_i(\textbf{r}+\textbf{v}_i \Delta t,t+\Delta t) = f_i(\textbf{r},t)+\Omega_i.
\end{equation}
The BGK collision operator is
\begin{equation}
    \Omega_i = \sum_j \Lambda_{ij} [f_j^\mathrm{eq}-f_j(\textbf{r},t)],
\end{equation}
where the $f_i$ are continuous densities associated with a lattice velocity $\textbf{v}_i$ that represent an expectation value for the number of particles moving from lattice site $\textbf{r}-\textbf{v}_i\Delta t$ to lattice site $\textbf{r}$ at time $t$. The BGK collision term redistributes those densities and relaxes them towards an imposed local equilibrium distribution $f_i^\mathrm{eq}$. In the simplest case, the relaxation matrix $\Lambda_{ij}$ has a single relaxation time $\Lambda_{ij}=(1/\tau) \delta_{ij}$, where $\tau=1$ implies that local equilibrium is reached in one time step.  For these methods the viscosity is 
\begin{equation}
  \nu=(\tau-0.5)/3,
  \label{eqn:LBnu}
\end{equation}
where the offset of $0.5$ is a result of recombining terms from the Taylor expansion of the occupation probabilities with the terms obtained from the continuous Boltzmann equation. A general $\Lambda_{ij}$ leads to multiple relaxation times, which is unimportant for this letter, since only one relaxation time turns out to be relevant for simple shear.

BGK lattice Boltzmann methods can no longer be justified as ensemble averages of the Boolean lattice gas models, and it became necessary to consider an alternative way of deriving the lattice Boltzmann method\,\cite{he1997theory}. This was achieved by deriving lattice Boltzmann directly as a discretization of the continuous Boltzmann equation. Decades later, it was realized that it is still possible to derive the BGK lattice Boltzmann methods from lattice gases with integer occupation numbers\,\cite{blommel_integer_2018}. 

However, any of these derivations require the relaxation time in Eq. (\ref{eqn:LBnu}) to be $\tau>1$, \textit{i.e.} the averaged collisions bring the distribution functions at most to local equilibrium. He, Chen and Doolen\,\cite{he1998novel} originally postulated that deriving lattice Boltzmann directly from the continuous Boltzmann equation could recover over\nobreakdash-relaxation. Later B\"osch and Karlin\,\cite{bosch2013exact} showed that it was only an uncontrolled approximation in their derivation that lead to this result, whereas an exact analysis showed that the regime of over\nobreakdash-relaxation is disconnected from the kinetic theory domain. Despite this apparent disconnect between over\nobreakdash-relaxation and physical theory, over\nobreakdash-relaxing the densities, \textit{i.e.} using $0.5\leq\tau<1$, is extremely useful in obtaining lower viscosities, and is frequently used in practical applications.

Clearly, lattice Boltzmann methods with over-relaxation can no longer be related to lattice gas methods by a statistical average, since the local collisions can only achieve equilibrium, but never over\nobreakdash-relax. Deriving lattice Boltzmann methods directly from discretizations of the Boltzmann equation equally fails to justify the usage of over\nobreakdash-relaxation as was shown by B\"osch and Karlin\,\cite{bosch2013exact}.

One could argue that is not important if lattice Boltzmann methods can be connected to some underlying physical model, as long as the method performs well. We believe this to be a shortsighted view. Seeing the lattice Boltzmann method as just another way of discretizing the Navier\nobreakdash-Stokes equations misses the key ingredient allowing lattice Boltzmann to outperform classical Computational Fluid Dynamics (CFD) approaches in a number of areas. A stunning example is that lattice Boltzmann methods have all but displaced classical CFD from the modeling of automotive hydrodynamics and are making significant inroads in the aerospace industry \cite{HChenPriv}. The success of the lattice Boltzmann method must be firmly attributed to its grounding in some physical reality, and the inability of linking the frequently used over\nobreakdash-relaxation to a physical underpinning is a worrying shortcoming.

In this letter, we show how this shortcoming can be overcome by a novel way of deriving lattice Boltzmann methods using a direct mapping approach from an average over Molecular Dynamics (MD) simulations onto lattice Boltzmann method. This approach has its roots in the Molecular Dynamics Lattice Gas (MDLG) method, pioneered by Parsa \textit{et al.}\,\cite{parsa_lattice_2017}.

\begin{figure}
     \centering
     \subfloat[]{\label{subfig-2:mdlg_d2q9}{\includegraphics[width=0.5\columnwidth]{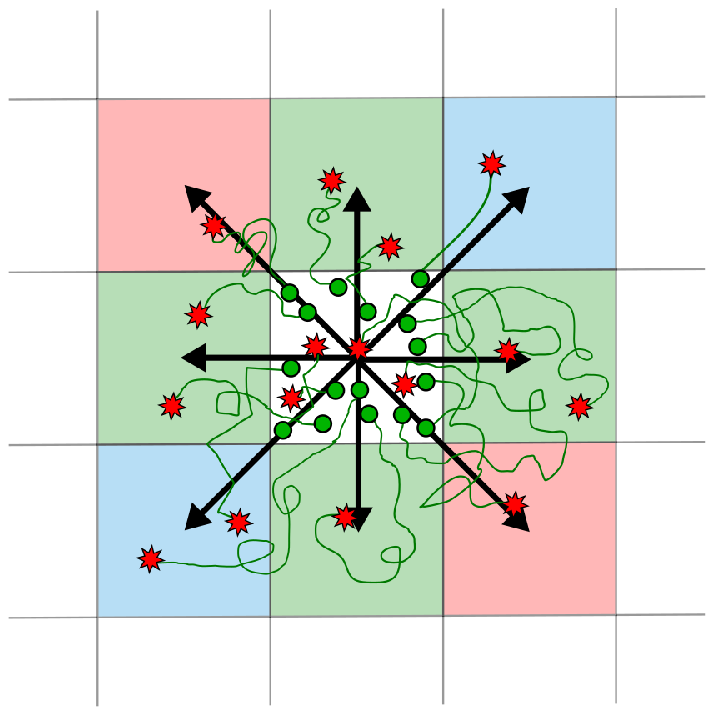}}}
     \subfloat[]{\label{subfig-2:mdlg_fi}{\includegraphics[width=0.5\columnwidth]{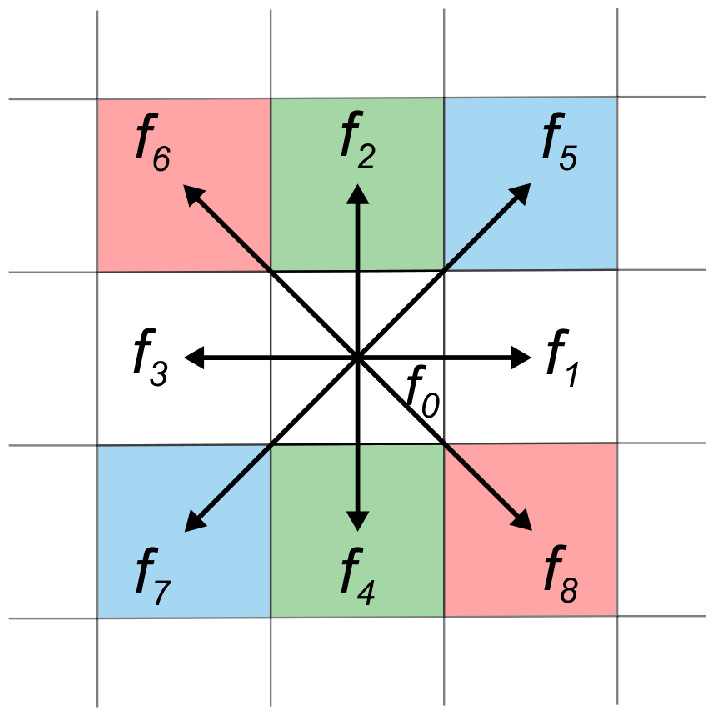}}}
     \caption{(Color online) (a) Sketch of the MDLG dynamics for D2Q9 lattice. A lattice is imposed onto the MD simulation domain and the movement of the particles is tracked from the central node using their MD trajectories. The circles (green) represent the particle position at time $t-\Delta t$ and the stars (red) are their respective positions at time $t$. The arrows (black) depict the lattice velocities. (b) Schematic representation of the D2Q9 lattice showing the numbering convention for $f_i$. The symmetries in the lattice are color-coded. }
\end{figure}

Briefly the MDLG method consists of overlaying a square lattice with lattice spacing $\Delta x$ onto an MD simulation. We define lattice displacements vectors $c_i$ connecting different lattice sites, using the index $i$ to enumerate the possible displacements. After fixing a time step $\Delta t$, we identify the number of particles that move from cell $\mathbf{r}-\mathbf{c}_i$ at time $t-\Delta t$ to lattice cell $\mathbf{r}$ at time $t$ with $n_i(\mathbf{r},t)$ lattice gas occupation number. This procedure maps the MD simulation onto a lattice gas\,\cite{parsa_lattice_2017} as shown in Fig.\,\ref{subfig-2:mdlg_d2q9}. The idea of the Molecular Dynamics Lattice Boltzmann (MDLB) is then to average over an ensemble of MD simulations of the same macroscopic state to obtain the lattice Boltzmann densities
\begin{equation}
    f_i(\mathbf{r},t) = \langle n_i(\mathbf{r},t)\rangle.
    \label{eq:fi_ni}
\end{equation}
Once we have $f_i(x,t)$, we can determine the lattice Boltzmann collision operator
\begin{equation}
    \Omega_i = f_i(\mathbf{r}+\mathbf{c}_i, t+\Delta t)-f_i(\mathbf{r},t).
    \label{eq:Omega}
\end{equation}
The focus of this letter are the properties of the MDLB collision operator and its ability to exhibit over-relaxation.

While the above described procedure is general and can, in principle, be applied to any flow, the numerical cost of averaging over a large number of MD simulations can be considerable. Instead, we investigate the simplest non\nobreakdash-equilibrium situation: a simple shear flow where the averaged velocities are given by
\begin{equation}
    u_x = \dot{\gamma} y;\;\;\;u_y=0,
\end{equation}
with $\dot{\gamma}$ being the shear rate, $\mathbf{r}=(x,y)^T$ is the position vector, and the density remains constant. Since, this flow is invariant under translation in the $x$-direction, and shifts in the $y$-direction can be related by a simple Galilean transformation to the $y=0$ position, all points are in this sense equivalent. Thus, we can average over all lattice points at all times, allowing for ample statistical averaging.

The MD simulations are executed using LAMMPS framework\,\cite{plimpton_fast_1995, noauthor_lammps_nodate} developed by Sandia National Laboratories. The system consists of particles interacting with the standard 6\nobreakdash-12 Lennard\nobreakdash-Jones (LJ) intermolecular potential. The particle mass $m$ and diameter $\sigma$ are set to one. Each simulation contains $N=99\ 856$ particles in a two-dimensional square with length L = 1000 LJ units referring to an area fraction of $\phi=0.078387$. The area fraction $\phi$ for circular LJ particles with van der Waals radius $r=\sigma/2$ is defined as the product of the particle surface area and the number of particles, divided by the square length of the simulation box. We initialised the simulations using homogeneously distributed particles with kinetic energy equal to 20 in LJ units, which corresponds to a dilute gas. We use the LAMMPS \texttt{nvt/sllod} thermostat to generate the desired non\nobreakdash-equilibrium dynamics. The lattice Boltzmann discretizations in time and space ($\Delta t$ and $\Delta x$) are fixed so that $\langle(\delta x)^2\rangle^\mathrm{eq}/(\Delta x)^2\approx \frac{1}{6}$,
where $\langle(\delta x)^2\rangle^\mathrm{eq}$ is the equilibrium mean\nobreakdash-squared displacement. This ratio ensures that the particle displacements are essentially limited to a neighborhood touching the central cell as shown in Fig.\,\ref{subfig-2:mdlg_d2q9}. This is referred to as an D2Q9 model since it resides in two dimensions and requires nine lattice velocities. We perform a wide range of simulations -- from simulations, where mean free time (\textit{i.e.} the time between collisions) is much larger than $\Delta t$ (ballistic regime) to simulations, where $\Delta t$ is much larger than the mean free time (diffusive regime). The data is collected after the simple shear has reached a steady state. For further information, please, refer to the supplemental material and the LAMMPS documentation\,\cite{noauthor_lammps_nodate}.

The symmetry of the simple shear flow puts significant constraints on the collision term $\Omega_i$ defined in Eq.\,(\ref{eq:Omega}). The degrees of freedom for the collision operator at the point $y=0$, where the mean velocity is zero, are sketched in Fig.\,\ref{subfig-2:mdlg_fi}. The point symmetry about the center of the lattice implies $f_2=f_4$, $f_5=f_7$, and $f_6=f_7$. Translational symmetry in the x\nobreakdash-direction implies that $f_0$, $f_1$, and $f_3$ are unchanged by the collision. Therefore, symmetry leaves only three independent values for an D2Q9 velocity set in the collision term $\Omega_i$, which is reduced to two because mass conservation adds the additional constraint $\sum_i \Omega_i=0$.
Therefore, the D2Q9 collision operator is determined by two terms that we choose as
\begin{align}
    \Omega^\alpha&=\Omega_6-\Omega_5+\Omega_8-\Omega_7,\\
    \Omega^\beta&=\Omega_2+\Omega_4,
\end{align}
with $\Omega^\alpha\gg\Omega^\beta$ for a simple shear. In this letter, we focus on the dominant collision contribution $\Omega^\alpha$. Now, we can define the moment before the collision as a function of the probability distribution function $f_i$
\begin{equation}
    M^\alpha = f_6-f_5+f_8-f_7,
    \label{Ma}
\end{equation}
and the moment after the collision
\begin{equation}
M^{\alpha,*}=M^\alpha+\Omega^\alpha.
\end{equation}
In equilibrium, we have $M^{\alpha,\textrm{eq}}=0$ due to symmetry.
\begin{figure}
     \centering
     \includegraphics[width=\columnwidth]{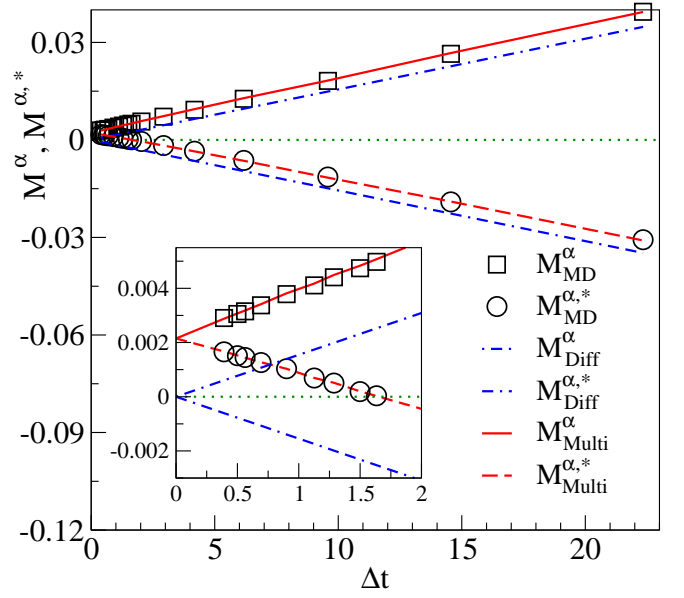}
     \caption{(Color online) The pre- and post-collision moments ($M^\alpha$ and $M^{\alpha, *}$) are shown as a function of $\Delta t$. The sign change between $M^\alpha$ and $M^{\alpha, *}$ indicates the transition from under- to over-relaxation. Three sets of data are shown: the measured from MD depicted as symbols (black); analytical solution using the multivariate Gaussian with diffusive moments depicted as dash-dotted lines (blue); analytical solution using the multivariate Gaussian with measured moments depicted as full and dashed lines (red). The zoomed plot shows the sign change of $M^{\alpha,*}_\mathrm{MD}$ and $M^{\alpha,*}_\mathrm{Multi}$ at $\Delta t\approx1.6$.}
    \label{fig:displ_moments}
\end{figure}
The signature of over\nobreakdash-relaxation is, therefore, a sign change between $M^\alpha$ and $M^{\alpha,*}$. The measured values of these quantities are shown as symbols in Fig.\,\ref{fig:displ_moments} as a function of $\Delta t$. For small $\Delta t$ both the $M^\alpha_\mathrm{MD}$ and $M^{\alpha,*}_\mathrm{MD}$ are positive, but $M^{\alpha,*}_\mathrm{MD}$ changes sign for $\Delta t\gtrapprox1.6$. Hence, the MDLB procedure predicts that for larger coarse\nobreakdash-graining the relaxation towards equilibrium is replaced by an over\nobreakdash-relaxation.

\begin{figure}
     \centering
     \includegraphics[width=\columnwidth]{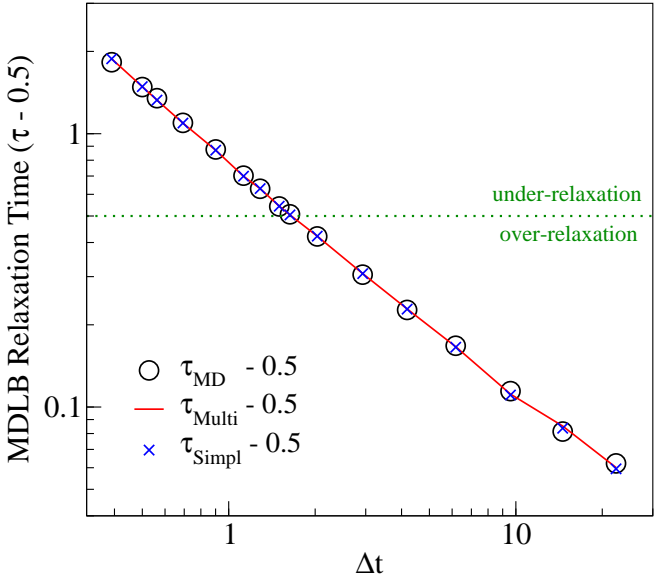}
     \caption{(Color online) The relaxation time $(\tau - 0.5)$ as a function of $\Delta t$ (logarithmic scale). Three sets of data are shown: measured from MD simulation depicted as circles (black); multivariate Gaussian with the MD depicted as a line (red); simplified multivariate Gaussian with $\langle(\delta x)^2\rangle = \langle(\delta y)^2\rangle = \langle(\delta x)^2\rangle^\mathrm{eq}$ depicted as crosses (blue). For $(\tau-0.5)$ above the dotted line (green), the collision operator under-relaxes, while for $(\tau-0.5)$ below that line, the collision over-relaxes.}
    \label{fig:relaxation_time}
\end{figure}

In terms of the relaxation time $\tau$, we have
\begin{equation}
    M^{\alpha,*} = M^\alpha + \frac{1}{\tau} (M^{\alpha,eq}-M^\alpha),
    \label{Ma*}
\end{equation}
with $M^{\alpha,eq}=0$, the relaxation time can be expressed as
\begin{equation}
    \tau = \frac{M^\alpha}{M^\alpha-M^{\alpha*}}.
\end{equation}
In Fig.\,\ref{fig:relaxation_time}, we show $(\tau - 0.5)$ as a function of $\Delta t$.

The remainder of this letter focuses on the origin of the observed over\nobreakdash-relaxation. In the MDLG coarse\nobreakdash-graining, $f_i$ can be expressed in terms of the one\nobreakdash-particle displacement function $P(\mathbf{r},\mathbf{\delta r})$\,\cite{parsa_lattice_2017} 
\begin{align}
    f_i(\xi,t) = \int_x P(\mathbf{r},\mathbf{\delta r})\Delta_{\mathbf{\xi}-\mathbf{c}_i}(\mathbf{r}-\delta \mathbf{r})\Delta_\xi(\mathbf{r}) d\mathbf{r},
    \label{eqn:fi}
\end{align}
where $\Delta_\xi(\mathbf{r})$ is one, if $\mathbf{r}$ resides in the lattice site $\mathbf{\xi}$ and zero otherwise, and $\mathbf{c}_i=\mathbf{v}_i\Delta t$ is the lattice displacement. This reduces the problem of finding $f_i$ to the problem of finding the one\nobreakdash-particle displacement function, which can be very challenging for arbitrary flows. In the diffusive limit, \textit{i.e.} when the mean free path is small and particles effectively undergo Brownian motion, an analytical solution exist\,\cite{elrick1962,van1977,foister1980}. In this case, the one\nobreakdash-particle displacement distribution function is given by a multivariate Gaussian probability distribution
\begin{equation}
\begin{split}
     &\hspace*{-0.25cm}P(x,y,\delta x, \delta y) = \frac{\sqrt{-\frac{\sigma_x}{\sigma_{xy}^2}+\frac{4}{\sigma_y}}}{2\pi\sqrt{\sigma_x}}\\&\hspace{-0.35cm}\times\exp{\left(-\frac{(\delta x-y\dot{\gamma} \Delta t)^2}{\sigma_x}-\frac{(\delta x- y\dot{\gamma} \Delta t) \delta y}{\sigma_{xy}}-\frac{(\delta y)^2}{\sigma_y}\right)}
     \label{eq:multivariate_pdf}
\end{split}
\end{equation}
with the moments
\begin{equation}
  \begin{split}
    \sigma_x &= \langle (\delta x)^2\rangle^{\mathrm{eq}} (1+\frac{\dot{\gamma}^2\Delta t^2}{3}),\\
    \sigma_{xy} &= \frac{\langle (\delta x)^2\rangle^{\mathrm{eq}} \Delta t\dot{\gamma}}{2},\\
    \sigma_y &= \langle (\delta x)^2\rangle^{\mathrm{eq}},
    \end{split}
    \label{eqn:sigmadiff}
\end{equation}
where $\langle (\delta x)^2\rangle^{\mathrm{eq}}$ is the measured equilibrium mean\nobreakdash-squared displacement as defined in \cite{pachalieva_2020, pachalieva_2021}. Note that a Galilean transformation is applied to the $x$-displacements that are at $y\neq 0$. Using Eqs.\,(\ref{eqn:fi})-(\ref{eqn:sigmadiff}) we calculate $f_i$ and obtain $M^\alpha_\mathrm{Diff}$, and $M^{\alpha,*}_\mathrm{Diff}$, which are shown as dash-dotted lines (blue) in Fig.\,\ref{fig:displ_moments}. The trend is very similar to the MD measurements but the results obtained using the diffusive moments are offset by a constant. The analytical result is entirely symmetric around the origin, leading to a relaxation time of $0.5$ for all time steps.

If we instead use a multivariate Gaussian with moments measured in the MD simulation
\begin{equation}
    \sigma_x = \langle (\delta x)^2\rangle,\;\;\;\;\;\sigma_{xy}=\langle\delta x \delta y \rangle, \;\;\;\;\;\sigma_y=\langle (\delta y)^2\rangle,
    \label{eq:md_moments}
\end{equation}
we obtain the predictions for $M^\alpha_\mathrm{MD}$, shown as solid line (red), and  $M^{\alpha,*}_\mathrm{MD}$, shown as dashed line (red), in Fig.\,\ref{fig:displ_moments}. They are in excellent agreement with the measured values. In Fig.\,\ref{fig:relaxation_time}, we show that the resulting relaxation time ($\tau_\mathrm{Multi}-0.5$) is likewise in excellent agreement with the measurement ($\tau_\mathrm{MD}-0.5$).

\begin{figure}
    \centering
    \includegraphics[width=\columnwidth]{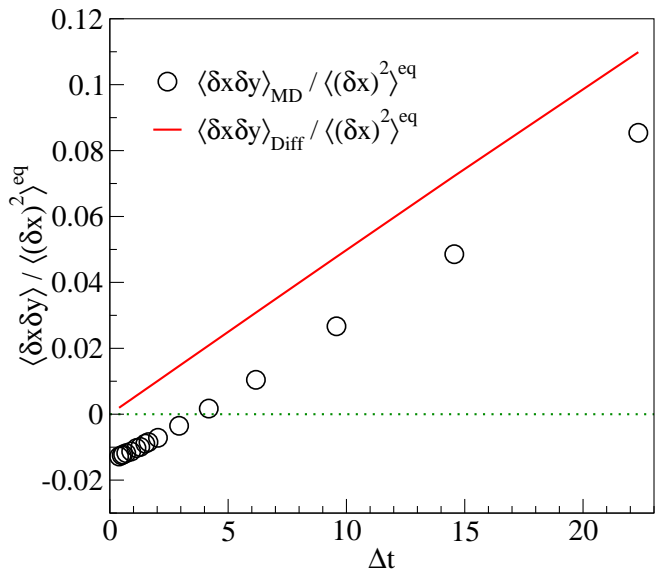}
    \caption{(Color online) The off-diagonal moment $\langle\delta x\delta y\rangle$ bears the main change from under- to over-relaxation, shown here as the moment is normalized by the equilibrium mean-squared displacement $\langle(\Delta x)^2\rangle^\mathrm{eq}$. The circles (black) depict $\langle\delta x\delta y\rangle_\mathrm{MD}$ measured from the MD simulation. The line (red) depicts $\langle\delta x\delta y\rangle_\mathrm{Diff}$ obtained using the multivariate Gaussian with diffusive moments. The sign change of $\langle\delta x\delta y\rangle$ is a key feature of the transition from under- to over-relaxation, however, it is not the only factor since the over-relaxation occurs at $\Delta t\approx 1.6$ and the sign change happens at $\Delta t\approx 4$. }
    \label{fig:dxdy_dxdx}
\end{figure}

To understand the physical origin of the transition from under- to over\nobreakdash-relaxation let us make a few observations: for the modest shear considered here with $(\dot{\gamma}\Delta t)^2\ll 3$ in Eq.\,(\ref{eqn:sigmadiff}), we have $\langle (\delta x)^2\rangle \approx \langle (\delta y)^2\rangle$ and both are approximately given by the equilibrium mean\nobreakdash-squared displacement $\langle(\delta x)^2\rangle^\mathrm{eq}$. The key change occurs in the off\nobreakdash-diagonal moment $\langle \delta x \delta y\rangle$. In Fig.\,\ref{fig:dxdy_dxdx}, we show $\langle \delta x \delta y\rangle/\langle (\delta x)^2\rangle^\mathrm{eq}$ as a function of $\Delta t$. We depict the off\nobreakdash-diagonal moment measured from the MD simulation with symbols (black) and the one calculated using the diffusive moments in Eq.\,(\ref{eqn:sigmadiff}) with a line (red).
The off\nobreakdash-diagonal moment $\langle \delta x \delta y\rangle_\mathrm{MD}$ changes sign at $\Delta t\approx4$ and otherwise behaves similar to the diffusive theory, albeit with an offset. The qualitative behavior in the diffusive case is straight forward: as particles diffuse into the positive $y$-direction they get carried away with the flow, and obtain an additional positive $x$-displacement leading to a positive correlation between $x$- and $y$-displacements. This means that any memory is quickly lost in frequent collisions. In the ballistic case, however, collisions are rare, and particles carry a memory of their history over larger distances. In particular, particles that move into the positive $y$-direction will typically have last collided at a position with negative $y$. In these regions, the average velocity is negative, so these particles will carry the average negative $x$-velocity prevalent in the region of their last collision to the regions of larger $y$. This leads to an anti-correlation between the $x$- and $y$-displacement. As we are looking at larger $\Delta t$, collisions become more frequent, and eventually the diffusive behavior becomes dominant. 

The predication of the relaxation time $(\tau_\mathrm{Simpl}-0.5)$ in Fig.\,\ref{fig:relaxation_time}, is calculated using a simple model with $\sigma_x=\sigma_y=\langle(\delta x)^2\rangle^{\mathrm{eq}}$ and the measured off\nobreakdash-diagonal moment $\langle \delta x \delta y\rangle$. We see that this simple model is also in excellent agreement with the measured values, showing that the off\nobreakdash-diagonal moment is indeed responsible for the transition from under- to over\nobreakdash-relaxation. Note, however, that it is not simply the sign change that determines this transition as the sign change occurs at $\Delta t\approx 4$ whereas the transition from under- to over\nobreakdash-relaxation occurs at $\Delta t\approx 1.6$.

In conclusion, in this letter we have shown that a lattice Boltzmann collision operator can be directly derived from one\nobreakdash-particle displacement probability distribution, which can be obtained from an MD simulation. This approach shows that such lattice Boltzmann collision operators naturally transition from under- to over\nobreakdash-relaxation. Thus, the over\nobreakdash-relaxation in lattice Boltzmann methods can be derived from first principles and is a consequence of the coarse\nobreakdash-grained representation of a lattice Boltzmann method.
%%%%%%%%%%%%%%%%%%%%%%%%%%%%%%%%%%%%%%%%%%%%%%%%%%%%%%%%%%%%%%%%%%%%%%%%%%%%%%%%%%%%
\bibliography{references_05_06}

\end{document}